\documentclass[nohyper,12pt,letterpaper]{JHEP}
\usepackage{axodraw}

\newfont{\frak}{eufm10 scaled 1200}

\newfont{\Bbb}{msbm10 scaled 1200}     
\newcommand{\mathbb}[1]{\mbox{\Bbb #1}}
\DeclareSymbolFont{AMSa}{U}{msa}{m}{n}
\DeclareSymbolFont{AMSb}{U}{msb}{m}{n}
\let\Box\relax
\DeclareMathSymbol{\Box}{\mathord}{AMSa}{"03}

\title{Breaking SUSY on the Horizon}

\author{T. Banks\thanks{On leave from NHETC, Rutgers U., Piscataway, NJ}\\
  Department of Physics and SCIPP\\
  University of California, Santa Cruz, CA 95064\\
  E-mail: \email{banks@scipp.ucsc.edu}}

\abstract{I present a heuristic calculation of the critical exponent
relating the gravitino mass to the cosmological constant in a de
Sitter universe.  The ingredients for the calculation are the
area law for entropy, an R symmetry of the low energy effective
Lagrangian, and a crude picture of the degenerate levels of the
cosmological horizon. }


\received{???????? ?st, 2000} \accepted{???????? ?th, 2000}

\preprint{\hepth{0206117}\\RUNHETC-2002-16\\SCIPP-02/41}
\begin{document}

\section{\bf Introduction }

In \cite{tbfolly} I proposed that the breaking of Supersymmetry
(SUSY) in the world we observe is correlated with a nonzero value
of the cosmological constant.  Crucial elements of this
conjecture were the claim that Poincare invariant theories of
gravity had to be exactly supersymmetric, and the claim that the
cosmological constant is an input parameter, determined by the
finite number of quantum states necessary to describe the
universe\footnote{One can play with the idea of a meta-theory in
which universes with different numbers of states are generated by
some random or deterministic process, and the number
characterizing our world is picked out by anthropic or number
theoretic criteria.  Such a theory could never be subjected to
experimental test, and so appears somewhat futile.}.  It follows
that the gravitino mass is a function of $\Lambda$, vanishing as
$\Lambda \rightarrow 0$ and the number of states goes to infinity.
As in any critical phenomenon, one may expect that classical
estimates of the critical exponents are not correct.  Thus I
proposed that the classical formula $m_{3/2} \sim \Lambda^{(1/2)}$
(in Planck units) might be replaced by $m_{3/2} \sim
\Lambda^{(1/4)}$ in a correct quantum mechanical calculation. The
latter formula has been known for years to predict TeV scale
superpartners when the cosmological constant is near the current
observational bounds.

Unfortunately, I was not able to come up with even a crude
argument for the validity of this conjecture.  Indeed, in a
recent note on the phenomenology of cosmological SUSY breaking
(CSB) \cite{scpheno}, I entertained the hypothesis that various
terms in the low energy effective Lagrangian scaled with different
powers of $\Lambda$.  I am happy to report that this state of
affairs has changed.  Where there was nothing, there is now a
waving hand.  That is, there is a set of plausible sounding
arguments about the interaction of particles with the
cosmological horizon that reproduces the critical exponent $1/4$
for the gravitino mass.  I do not pretend that these arguments
are definitive, but I do hope that they are approximately correct.

In my initial thinking about this problem, I suggested that
"Feynman Diagrams'' describing virtual black hole production and
decay were responsible for the anomalous relation between the
gravitino mass and the cosmological constant.  I soon realized
that this was unlikely to make sense.  Although production of
black holes by high energy few particle collisions has
probability of order one, the probability that the decay products
of a large black hole will reassemble themselves in spacetime so
that they can be absorbed by the particle that emitted the high
energy virtual lines , and contribute to its mass
renormalization, is infinitesimally small.  While this argument
involves an extrapolation between onshell and offshell processes,
it convinced me that black holes were not the answer.
Simultaneously, I realized that most of the states in dS space,
could not be described as black holes in a single observer's
horizon volume.  Rather they should be thought of as black holes
sprinkled among many different static horizon volumes, causally
disconnected from each other.  According to cosmological
complementarity \cite{bfs}, a single observer sees these as
states near his cosmological horizon.  This observation led to
the paper which follows.

The calculation I will present is very heuristic and unconventional and
one might be led to ask why more conventional methods of quantum field
theory in curved spacetime do not lead to hints of this behavior.
In fact, no calculations have been done in the conventional framework,
and there are many indications that any such calculation would suffer
from a variety of divergences.  I will outline some of the problems in
an appendix.

\section{The Low Energy Effective Lagrangian}

According to the conjecture of \cite{tbfolly}\cite{wf},
Asymptotically dS (AsdS) spaces have a finite number of quantum
states.  In a universe with a finite number of states, there can
never be precisely defined observables, because any exact
measurement presupposes the existence of an infinite classical
measuring apparatus.  With a finite number of states (all of
which have at least mutual gravitational interactions) there is
no way to make a precise separation between observed system and
measuring apparatus, nor any possibility of exactly neglecting the
quantum nature of the measuring apparatus.  Thus there should be
no mathematically defined observables in dS spacetime.

Nonetheless, we know (though we do not yet know why or how) that
when the cosmological constant is small, there should be an
approximate notion of scattering matrix and (at low enough
energy) of an effective Lagrangian.   The hypotheses of
\cite{tbfolly} make this more precise.  The cosmological constant
is a tunable parameter, and in the limit that it vanishes there
is a SUSic theory of quantum gravity in asymptotically flat
spacetime.  This theory has a well defined S-matrix.  There
should be an object for finite $\Lambda$ which converges to this
S-matrix.  Indeed, by analogy to critical phenomena, one might
expect that there are a plethora of different unitary operators
in the finite $\Lambda$ Hilbert space, which all converge to the
same S-matrix.  However, again by analogy to critical phenomena,
we might expect that several terms in the asymptotic expansion of
the S matrix around $\Lambda = 0$ are {\it universal}.  I believe
that it is in these universal terms that the "physics of AsdS
space" lies.  Everything else will be ambiguous.

I expect that the validity of this asymptotic expansion is highly
nonuniform, both in energy and particle number.  This is a
consequence of the postulated finite number of states of the AsdS
universe.   The best convergence is to be expected for low
energy, localizable processes, which do not explore most of the
spacetime.   These are the processes described by the low energy
effective Lagrangian.  The gravitino mass is the coefficient of a
term in this Lagrangian, and has no more exact definition.

In the $\Lambda \rightarrow 0 $ limit the effective Lagrangian
should become SUSic and of course have vanishing cosmological
constant.  This indicates \cite{scpheno} that a complex
(discrete) R symmetry is also restored in this limit.  In
\cite{scpheno} I argued that the limiting theory had to be a four
dimensional $N=1$ SUGRA with a massless chiral or vector multiplet
which will be eaten by the gravitino when $\Lambda$ is turned
on.  From the point of view of the effective Lagrangian, SUSY
breaking should be spontaneous and triggered by explicit R
breaking terms, all of which vanish as some power of $\Lambda$.
Among these is a constant in the superpotential, which guarantees
that the cosmological constant take on its fixed input value.
{}From the low energy point of view this looks like fine tuning.
{}From a fundamental point of view it is merely a device for
assuring that the low energy theory describes a system with the
correct number of states (once one implements the
Bekenstein-Hawking bound).

The reason that R breaking is explicit while SUSY breaking must
be spontaneous is that the R symmetry is discrete, while SUSY is
an infinitesimal local gauge symmetry.  Explicit SUSY breaking
can be made to look spontaneous by doing a local SUSY
transformation and declaring that the local SUSY parameter is a
Goldstino field.   The possibility of restoring SUSY by tuning
the cosmological constant to zero implies that the Goldstino must
come from a standard linearly realized SUSY multiplet, which
appears in the low energy Lagrangian.  No such arguments are
available for the discrete R symmetry.   The picture of low
energy SUSY breaking which thus emerges consists of a SUSic, R
symmetric theory, in which SUSY is spontaneously broken once R
violating terms are added to the Lagrangian.  The R violation
should be attributed to interaction of the local degrees of
freedom with the cosmological horizon.

The purpose of this paper is to estimate the size of the R
breaking terms in the low energy effective Lagrangian.  In
\cite{scpheno} I also had to invoke Fayet-Iliopoulos (FI) D terms
for some $U(1)$ groups .  I have since learned from Ann Nelson
that such terms can be generated by nonperturbative physics in
low energy gauge theory\cite{anetal} .  This physics in turn
depends on the existence of certain terms in the superpotential;
terms which could be forbidden by an R symmetry.  In addition, I
have found models not considered in \cite{scpheno}, where
dynamical SUSY breaking is triggered by the addition of R
violating terms to an otherwise SUSic theory. I will therefore
assume that our task is just to estimate the breaking of R
symmetry by the horizon. As we will see, this is dominated by the
lightest R-charged particle in the bulk.  In many models of low
energy physics, this will be the gravitino, and I will assume that
this is the case.

\section{Horizonal Breaking of R Symmetry}

The basic process by which the horizon can effect the low energy
effective Lagrangian is described by Feynman diagrams like that
of Fig. 1.  A gravitino line emerges from a vertex localized near
the origin of some static coordinate in dS space, propagates to
the horizon, and after interacting with the degrees of freedom
there, returns to the vertex.  The dominance of diagrams with
gravitinos is a consequence of our attempt to calculate R
violating vertices and our assumption that the gravitino is the
lightest R charged particle.  The dominance of diagrams with a
single gravitino propagating to the horizon will become evident
below.

In field theory, the effective Lagrangian induced by a diagram
like Fig. 1 will have a factor\footnote{All of the calculations of
this section are done in four dimensions.}

$$\delta{\cal{L}} \sim e^{- 2 m_{3/2} R} R^{-4},$$
\noindent
where $R$ is the spacelike distance to the horizon.  This factor
comes from the two propagators and an integral over the point
where the gravitino lines touch the horizon.  The gravitino is
assumed massive because we know that SUSY is broken in dS space.

\vfill\eject
\vskip1.5in
\begin{center}

\begin{picture}(400,100)(-100,-40)

\CArc(100,20)(100,0,360)

\ArrowLine(80,20)(100,120)

\ArrowLine(120,20)(100,120)

\CArc(100,20)(19.8,180,360)

\Line(80,20)(120,20)

\Line(100,.2)(100,-19.8)

\Line(86.1,6.1)(72.3,-7.7)

\Line(113.9,6.1)(127.7,-7.7)



\end{picture}\\
\vskip1in
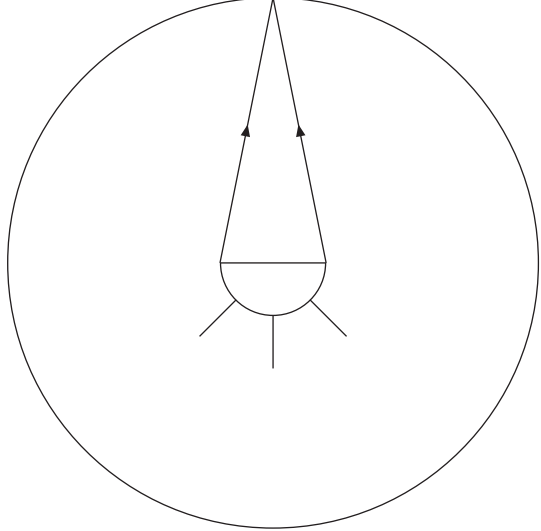
 {\sl Fig. 1 Effective Vertex Induced by Gravitino Exchange
With the Horizon} \end{center}

\vskip.3in

Our hope to overcome this field theoretic suppression comes from
the fact that the horizon has $e^{(RM_P)^2 \over 4}$ states.   We
thus want to estimate how many of these states the gravitino line
interacts with.  Since the entropy of the horizon is extensive,
this is, crudely, the amount of horizon area the gravitino sees.
The thermal nature of Hawking radiation from the horizon suggests
that the gravitino interacts in much the same way, with most of
the horizon states it comes in contact with.  Since the gravitino
is massive and the horizon a null surface, it can only propagate
along the null surface for a proper time of order $1/m_{3/2}$.

During its contact with the horizon the gravitino is interacting
with a Planck density of degrees of freedom.  Rather than free
propagation, we should imagine that it performs a random walk
with Planck length step over the surface of the horizon, covering
a distance of order $m_{3/2}^{- {1\over 2}}$ and an area of order
$1/m_{3/2}$.

We now make our major assumption, which is that the diagram of
Fig. 1 gets a coherent contribution from a number of states of
order $e^{a \over m_{3/2}}$ that the gravitino encounters as it
wanders over the horizon.   For small gravitino mass, this might
be an extremely tiny fraction of the total number of states
localized in the area of the horizon explored by the gravitino.
Thus, to exponential accuracy, the contributions to the R
breaking part of the low energy effective Lagrangian are of order
$$\delta {\cal L} \sim e^{- 2 m_{3/2} R + {a M_P\over m_{3/2}}}$$

We imagine the calculation of this diagram to be part of a self
consistent calculation of the gravitino mass, in the spirit of
Nambu and Jona-Lasinio \cite{NJL}.  That is, the interaction with
the horizon of a gravitino of a certain mass causes R breaking,
which gives rise to SUSY breaking and a gravitino mass. If
$m_{3/2}/M_P >  ({2RM_P \over a})^{- {1\over 2}}$ then our
calculation gives an $R$ breaking effective Lagrangian which falls
exponentially with $RM_P$ .  Thus, the horizon contribution is
totally negligible.  On the other hand, we know of no other
contribution to the mass which is this large, for large $RM_P$.
Thus masses in Planck units greater than $ ({2R M_P \over a})^{-
{1\over 2}}$ are not self consistent.

If $m_{3/2}/M_P < ({2RM_P \over a})^{- {1\over 2}}$, the equation
predicts an exponentially growing breaking of $R$ symmetry, and a
correspondingly huge gravitino mass, so again the assumption is
inconsistent.  Notice that this, in particular, rules out the
classical formula , $m_{3/2} \sim R^{-1}$.   The only self
consistent formula, to leading order in $RM_P$, is $m_{3/2} = M_P
({2RM_P \over a})^{- {1\over 2}}$.   Taking $R$ of order the
Hubble radius of the observable universe, we get a gravitino mass
of order $10^{- 11.5} GeV$.  This corresponds to a scale for the
splitting in nongravitational SUSY multiplets of order $5-6$ TeV.
This is the scaling of the gravitino mass conjectured in
\cite{tbfolly}.

It may appear that our solution for $m_{3/2}$ is not consistent
at the power law order in $RM_P$.  That is to say, if the $R$
dependence of $\delta\cal{L}$ is given by the above equation, it
does not give rise to a gravitino mass of order $\Lambda^{1/4} =
R^{-1\over 2}$However, corrections to the parameter $a$ of the
form $\delta a \sim b {\rm ln (RM_P) \over (RM_P)^{1/2}}$ can
remedy this difficulty. Alternatively, (or in addition)
nonleading, logarithmic terms in the self consistent formula for
$m_{3/2}$ for fixed $a$ can have the same effect.   Notice that
although at present we have no way of estimating such corrections,
self consistency requires them to be present in precisely the
right amounts.  The dominant exponential terms in the gravitino
mass relation are self consistent only for $m_{3/2}\sim
\Lambda^{1/4}$.

It should now also be clear why diagrams with more than one
gravitino line propagating to the horizon, or with any heavier
particle replacing the gravitino, are subdominant.  These have a
larger negative term in the exponential, but no larger
enhancement from the number of states.

Our calculation clearly rests squarely on the assumption that we
can get a contribution of order $e^A$ from interacting with an
area $A$ of a system.  This sounds peculiar when heard with the
ears of local field theory.  If, in the spirit of the Membrane
paradigm, we modeled the physics of the horizon by a cutoff field
theory we would again find $e^{A \sigma}$ states in an area $A$,
where $\sigma$ is the entropy density of the field theory.  Yet
the interactions of a probe concentrated in an area $A$ would be
expected to renormalize the effective action of the probe by
terms of order $A$.  This follows from general clustering and
locality arguments in field theory.

However, there is another way that such a field theoretic model
fails to capture the physics of horizons.  Consider the case of a
black hole.  A field theory model would predict an energy density
as well as an entropy density.   The total energy of the black
hole would then be of order its area, much larger than its mass.

A somewhat better model of a horizon may be obtained by
considering a system of fermions on a two sphere, coupled to an
external $U(1)$ gauge field\footnote{I would like to thank O.
Narayan for discussions about quantum Hall systems.}.  The field
configuration on the sphere is that produced by a magnetic
monopole of very large charge.   All fermions are in the lowest
Landau level and we tune the magnetic charge so that this is
completely full.   We can choose linear combinations of the
single particle wave functions so that each fermion is localized
in a quantum of area on the sphere.  Now imagine giving each
fermion a two valued ''isospin" quantum number, on which neither
the horizon's Hamiltonian nor its coupling to the external probe
depend.  The probe is coupled to the position coordinates of the
fermions, in a local manner. The system has $2^A$ degenerate
states in area $A$ and the probes effective action will be
renormalized by an amount $\propto 2^A$.

\section{Discussion}

The handwaving nature of the arguments I have presented is
probably unavoidable at this stage of our understanding of
quantum gravity in de Sitter space.  To do better, we must first
construct a complete quantum model of de Sitter space, presumably
a quantum system with a finite number of states.  This
construction must recognize the approximate nature of any theory
in dS space.  There should not be precisely defined, gauge
invariant observables, corresponding to the fact that no precise
self-measurements can be carried out in a quantum system with a
finite number of states.  Rather we should look for
approximations to the Super Poincare Generators and S-matrix. We
should then understand how to identify an approximate notion of
low energy effective Lagrangian, which describes some of the
physics of the full quantum S-matrix.  The latter in particular
is a difficult task. Even in Matrix Theory, and AdS/CFT, where
there is a precisely defined quantum theory, we can only find the
low energy effective Lagrangian by computing the S-matrix and
taking limits\footnote{In perturbative string theory we can also
derive the Lagrangian from the world sheet renormalization group,
an intriguing hint, which has not had any echoes in
nonperturbative physics.}.

Further development of the kind of mongrel argument used in this
note, in which properties of the low energy effective Lagrangian
and the horizon, are used as separate constructs which have to
fit into a self consistent picture, depends on refinement of our
understanding of horizons.  It may be that this can be achieved
by studying Schwarzchild black holes, and assuming that the local
properties of dS horizons are similar.  Black holes are objects
in asymptotically flat space and precise mathematical questions
about their properties can be formulated.  It is intriguing that
we have already found that a picture of horizon dynamics in terms
of a cutoff local quantum field theory on the horizon is
inconsistent both with the large breaking of SUSY conjectured in
\cite{tbfolly} and with the entropy/mass relation of the black
hole.   In this context, it is worth pointing out that for those
near extremal black holes where a field theoretic counting of
entropy is successful, the field theory does not live on the
horizon, and some of the horizon coordinates are quantum
operators. This suggests, as does the Landau level model of the
last section, that horizons are described by a noncommutative
geometry.

On a more phenomenological note, the present calculation sheds
some light on a possibility that I raised in \cite{scpheno}. I
suggested that different R breaking terms in the low energy
effect Lagrangian might scale with different powers of the
cosmological constant.   There is no hint of that possibility in
the calculation I presented.  That is, the powers of the dS
radius do not depend at all on the external legs of the Feynman
diagram, which would distinguish between different R breaking
operators.  On the other hand, if the FI D term is generated, as
in \cite{anetal} from low energy dynamics , which is itself
triggered by the existence of R breaking terms, the FI D term
might depend on both the explicit R breaking scale and the
dynamical scale of a low energy SUSY gauge theory. I will leave
the exploration of the latter question, as well as of alternative
models of low energy physics, to another paper.

It is worth pointing out that large effects of the type I have
calculated would not occur in FRW cosmologies which asymptote to
SUSic universes, at least if one follows the rules that I have
advocated here.  The holographic screen, analogous to the
cosmological horizon, for such a universe is future null infinity.
It is an infinite spacelike distance away from any finite point
on the worldline of a timelike observer.  The effect I have
calculated is larger than any which might have been found in
local physics, but it still vanishes as the spacelike distance to
the holographic screen goes to infinity.   SUSY breaking in such
spacetimes will be dominated by local physics.

A somewhat more puzzling situation is presented by a hypothetical
universe which stays in a dS phase for a very long time (60
gazillion years, to use technical language) but then asymptotes
to a SUSic state.  One can easily invent (fine tuned) models of
quintessence which have this property.  I think the answer here
is that the effective Lagrangian of a local timelike observer in
such a universe is time dependent.  It will initially exhibit
larger than normal SUSY breaking, and then become rapidly SUSic.
The puzzle that remains is how the transition is made and what a
convenient set of holographic screens for such a spacetime might
be.

 Finally, I want to respond to a question I have
been asked many times in the context of lectures on
\cite{tbfolly}: why doesn't CSB also imply large corrections to
other calculations in low energy effective field theory?   The
answer has been given before and does not really depend on the
calculation in this paper, but perhaps it obtains new force from
that calculation. The basic idea of CSB is that $\Lambda$ is a
variable parameter and that the $\Lambda = 0$ theory has a SUSic,
R symmetric low energy effective Lagrangian. All terms in this
Lagrangian obviously have a finite $\Lambda\rightarrow 0$ limit.
CSB is the theory of how the R violating terms (and perhaps an FI
D term that is induced by them) depend on $\Lambda$ in the flat
space limit.  It is calculating quantities that are parametrically
smaller than the SUSic, R symmetric terms in the Lagrangian.  Our
calculation did not lead to any terms, which diverge in the
limit.  Such behavior is not compatible with the self consistency
of the induced gravitino mass.  Thus, any corrections to the
preexisting terms in the Lagrangian take the form of small
(suppressed by a positive power of the cosmological constant)
corrections to their finite, SUSic, values.

\section{Appendix}

In order to discuss the question of SUSY breaking in dS space
we first have to realize dS space as a solution of a SUGRA Lagrangian.
This restricts us to minimal supergravity in 4 dimensions.
The simplest Lagrangian with a dS solution contains one chiral
supermultiplet in addition to the SUGRA multiplet, and is characterized
by a holomorphic superpotential $W(Z) $ and a Kahler Potential $K(Z,Z^*
)$.  In fact, away from zeroes and singularities of $W$, these can be
combined into a single function.  We will not make this combination
because we start out from the assumption of an R symmetric minimum where
$W$ vanishes.   The quantum field theory associated with this
Lagrangian is not renormalizable and must be supplemented with a cutoff
procedure, with a cutoff of order the Planck scale.  Ensuring that the
cutoff procedure is invariant under superdiffeomorphisms is a
complicated technical problem to which there is no known solution.
I will assume that this can be solved.

The presumed stationary point of $W$, with $W =0$ gives a
Minkowski spacetime in which can compute a gauge invariant
S-matrix, and extract from it, order by order in perturbation
theory, a gauge invariant effective action.  If we perturb the
superpotential by small explicit R breaking terms that lead to a
dS minimum with small cosmological constant, we would hope that,
at least to some order in the perturbation, the gauge invariant
effective action is still a valid physical quantity.  Indeed,
apart from the problem of superdiffeomorphism invariant
regulators, the standard background field action of DeWitt seems
to provide us with the required perturbatively gauge invariant
quantity, even for a dS background.  When I refer to particle
masses, I mean coefficients in this gauge invariant action. Note
that the action is globally dS invariant, so that even if one
insists that the global isometries are gauge transformations, the
effective action should still be meaningful.  In order to tune
the cosmological constant to be much smaller than the SUSY
breaking scale, $F = |DW|$ we must, generally, add an explicit R
breaking constant to $W$.

Now consider loop corrections to the effective action. These are
logarithmically UV divergent at one loop\footnote{It is often said that
in minimal four dimensional SUSY, the cosmological constant is
quadratically divergent.  It is hard to understand how this could
be compatible with the general structure of SUSic effective Lagrangians
for positive cosmological constant.  The positive term in the effective
potential is proportional to the square of the gravitino mass.
Thus, a divergent positive cosmological constant would imply a divergent
gravitino mass.  S.Thomas has informed me that if the calculation is
done with SUSic Pauli-Villars regulators, the one loop vacuum energy
is only logarithmically divergent.}.  Higher loop calculations will have
higher powers of the logarithm, and there is no small expansion
parameter.  In previous discussions of this problem, I have argued that
this shows UV sensitivity of the calculation of UV effects, and then
invoked the UV/IR
connection to claim that physics {\it above} the Planck scale will
renormalize the gravitino mass by amounts that depend on the
cosmological constant.

It now appears more probable that the signal for large breaking of
SUSY, within the framework of field theory, is the infamous
IR divergence problem of dS space\cite{IR}.  The transverse, traceless
part of the graviton propagator grows logarithmically at large
distances.  In an initial version of \cite{tbfolly} I considered
the possibility that these divergences might be the mechanism
responsible for the anomalous scaling behavior of the gravitino mass.
I rejected this mechanism because there was confusion in the literature
as to whether the IR divergences could appear in gauge invariant
physical quantities.   It was only later that I came to the conclusion
that there were no mathematically precise, gauge invariant quantities
in dS space.   IR divergences have certainly been shown to appear
in quantities that appear to be perturbatively gauge invariant.
I would speculate that the calculation of the gravitino mass
renormalization will similarly have IR problems.

It is not at all clear that any kind of resummation of these
divergences should give a finite answer, much less an answer that agrees
with the calculation of this paper.  Our calculation had an explicit
cutoff on the number of states, that is absent in QFT in dS space.
So the IR divergence of the naive low energy theory may just be an
indication that the true quantum theory of dS space is not well
approximated by field theory. At issue here is whether there is a sort
of duality between the description by a single static observer,
of UV processes localized near his cosmological horizon (this is the
description used in the foregoing paper), and an IR description of
the same physics using quantum field theory in the global dS space time.
None of our experience with black hole physics gives us guidance here,
since we do not yet have an adequate description of the physics of an
infalling observer.

If one believes in such a duality, he would be motivated to recover our
result by calculations in quantum field theory, that is, by trying to
resum the IR divergences that I believe would appear in a global
computation of the gravitino mass. On the other hand, for a given observer,
it may be that only the description of low energy processes localized
far from his/her cosmological horizon is well approximated by quantum
field theory.  At the moment, I do not know which of these two points of
view is correct, though the lack of a natural IR
cutoff in the field theory calculation suggests that one is unlikely to
reproduce the correct physics by simply resumming field theory diagrams.
Unfortunately, the calculation of loop corrections to the gravitino mass
in the perturbative approach to quantum gravity in dS space, is a
daunting exercise.  One may hope to compute the one loop contribution,
but a systematic analysis to all orders in perturbation theory, seems
difficult.

\acknowledgments
 This work supported in part by the U.S. Department of
Energy under grant DE-FG03-92ER40689.   I would like to thank
M.Dine, W. Fischler and O. Narayan for discussions.


\end{document}